# REPRESENTATION DES DEFAILLANCES DE CAUSE COMMUNE D'UN SYSTEME PROGRAMME DE GRANDE TAILLE PAR RESEAUX DE PETRI COLORES, TEMPORISES ET HIERARCHIQUES

Gilles Deleuze
EDF R&D
1 av. Gnal de Gaulle
92141 Clamart
gilles.deleuze@edf.fr

Nicolae Brinzei
CRAN - CNRS UMR 7039
Université de Lorraine
Avenue de la Forêt de Haye
54518 Vandoeuvre-lès-Nancy
nicolae.brinzei@univ-lorraine.fr

Nicolas Villaume
ENSEM
Avenue de la Forêt de Haye
54501 Vandœuvre-lès-Nancy

## Résumé

L'objet de cette étude est la représentation des Défaillances de Cause Commune (DCC) dans les systèmes programmés de grande taille. Le système considéré est représentatif du contrôle-commande d'une centrale nucléaire. Le modèle retenu de DCC est le modèle d'Atwood. En effet, il permet de représenter les défaillances indépendantes, les défaillances communes non létales pour certains composants du système et les défaillances communes létales pour tous. Pour cette étude, le modèle a été modifié pour « orienter » les DCC non létales sur certaines parties du système et ainsi prendre en compte les différentes origines possibles des DCC. La maintenance et les réparations sont prises en compte dans le modèle. La représentation du système doit donc être dynamique. La principale évaluation des résultats est probabiliste, la grandeur considérée est la probabilité de défaillance à la sollicitation (PFD). Une comparaison est faite entre l'estimateur de la PFD prenant en compte toutes les défaillances, la « PFD référence », et l'estimateur ne prenant en compte que les défaillances détectées, la « PFD visible ».

## Summary

The purpose of this study is the representation of Common Cause Failures (CCF) in large digital systems. The system under study is representative of a control system of a nuclear plant. The model for CCF is the generalized Atwood model. It can represent independent failures, CCF non-lethal for some system elements and CCF lethal to all. The Atwood model was modified to "direct" non-lethal DCC on certain parts of the system and take into account the different possible origins of DCC. Maintenance and repairs are taken into account in the model that is thus dynamic. The main evaluation results are probabilistic, the considered indicator is the probability of failure on demand (PFD). A comparison is made between the estimator of the PFD taking into account all the failures and the estimator taking into account only the detected failures.

## Contexte

Les Défaillances de Causes Communes (DCC) sont des défaillances en fonctionnement ou à la sollicitation pouvant affecter simultanément plusieurs composants d'un système et dues à une même cause. Les DCC affectent des groupes de composants identiques ou similaires, redondants, réalisant la même fonction et œuvrant dans des conditions comparables. Ainsi, les DCC contournent un des piliers de la sûreté de fonctionnement (SdF) : la redondance. Les DCC sont réduites par la diversité, mais jamais complètement. L'étude des Défaillances de Causes Communes (DCC) dans les systèmes programmés est un domaine actif de recherche dans les domaines du nucléaire, de la pétrochimie, de l'aérospatial…
Certains systèmes programmés, comme les systèmes de contrôle-commande du domaine nucléaire, sont de grande taille, fortement redondés et avec une logique de vote complexe. Ils sont sensibles à trois types de DCC distinctes : les DCC d'origine matérielle, les DCC d'origine logicielle (ou de conception) et les DCC d'origine humaine. Par ailleurs, le Retour d'Expérience (REX) de ces systèmes est difficilement exploitable car il y a peu d'événements.

Dans les grands systèmes redondants avec N éléments, un choc est supposé non létal (DCC partielle) lorsqu'il affecte k éléments parmi N avec $1 \leq k < N$. Un choc est létal lorsqu'il affecte tous les composants (DCC totale). De façon générale, un modèle simple de DCC, dit modèle du facteur $\beta$ [US-NRC, 1987], qui ne représente que les défaillances létales, ne donne pas satisfaction sur ce genre de système. En effet, ce modèle simple est adapté lorsque le système est constitué de quelques composants seulement. Lorsque le système est composé de plusieurs dizaines de composants identiques ou similaires, l'hypothèse de défaillance systématique de tous les composants, quand une CCF se produit, est très conservatrice.

De plus, les caractéristiques du système programmé étudié complexifient fortement les expressions analytiques pour obtenir les indicateurs de la SdF. La combinatoire introduite par la logique de vote, la présence de défaillances non détectées, la combinaison qui en résulte, de défaillances indépendantes et de DCC, en sont les principales raisons. Pour réduire la complexité des expressions analytiques, des simplifications sont opérées et sont à valider. L'estimation des indicateurs de la SdF avec un REX faible est une difficulté supplémentaire. Ainsi l'estimation de paramètres probabilistes représentant les DCC est difficile. Les données d'entrée utilisées, telles que les taux de défaillance et la proportion de défaillances détectées, sont également influentes. La simulation dynamique des DCC, pour estimer les paramètres DCC, est ainsi une solution pour valider les simplifications faites en absence de REX statistiquement significatif et estimer certaines incertitudes.

## Le modèle d'Atwood et son extension aux DCC orientées

### 1 Généralités

Dans cette partie, nous présentons le modèle d'Atwood. Le modèle d'Atwood fait partie de la famille des modèles de chocs, avec les modèles stochastiques, les modèles de chocs externes CLM et ECLM, les modèles de chocs internes. Le cadre théorique commun est le Modèle Résistance-Contrainte.

Ce modèle présente deux qualités principales :

- Un certain réalisme dans la représentation qualitative des stress pouvant causer des DCC. Le modèle d'Atwood, par sa structure, est adapté pour représenter la phénoménologie de nombreuses défaillances dépendantes affectant les systèmes programmés.
- Un nombre de paramètres réduit à 3, quelle que soit la taille du système.

Sa faiblesse principale réside dans la difficulté à corréler les valeurs de ses paramètres avec des observations empiriques. Ce modèle a été développé en deux étapes :

- Binomial Failure Rate (BFR) ou Modèle d'Atwood. Ce modèle a été proposé par [Atwood, 1980]. Il considère des défaillances indépendantes et des DCC partielles.
- Generalized Binomial Failure Rate (GBFRM) ou Modèle d'Atwood généralisé. Ce modèle, dit aussi Binomial Failure Rate Model (BFRM), est abondamment décrit dans la littérature depuis [Atwood, 1986]. Il introduit les chocs létaux dans le modèle BFR et prend en compte les défaillances de composants indépendants et les DCC dues aux chocs qui impactent l'ensemble ou seulement certains composants.

Seul le modèle dit « généralisé » est employé aujourd'hui. Les hypothèses qui sous-tendent les deux versions du modèle sont souvent vérifiées dans le cas du matériel électronique. Il existe des relations assez directes avec le modèle des facteurs partiels Beta(k,N).

## 2 Notations

| Notation | Notion équivalente pour un système en attente | Notion équivalente pour un système en fonctionnement continu |
|---|---|---|
| $Q_k^{(N)}$ Probabilité de défaillance d'un groupe spécifique de k éléments dans un groupe de N éléments, toutes causes confondues[1] | $\mathrm{Pfd}_k^{(N)}$ Probabilité de défaillance à la sollicitation d'un groupe G spécifique de k éléments dans un groupe de N éléments, toutes causes confondues | $\lambda_k^{(N)}$ Taux d'occurrence d'une défaillance d'un groupe G spécifique de k éléments dans un groupe de N éléments, toutes causes confondues |
| $Q_{TOT}$ Probabilité de défaillance d'un élément spécifique Ei, toutes causes confondues | $\mathrm{Pfd}_{,TOT}$ Probabilité de défaillance à la sollicitation d'un élément spécifique Ei, toutes causes confondues | $\lambda_{TOT}$ Taux d'occurrence d'une défaillance d'un élément spécifique Ei, toutes causes confondues |
| $Q_{IND}$ Probabilité de défaillance d'un élément spécifique Ei par défaillance indépendante[2] | $\mathrm{Pfd}_{,IND}$ Probabilité de défaillance à la sollicitation d'un élément spécifique Ei par défaillance indépendante | $\lambda_{IND}$ Taux d'occurrence d'une défaillance (toutes causes confondues) d'un élément spécifique Ei par défaillance indépendante |

**Table 1.** Notations concernant les probabilités conditionnelles et les taux de défaillance

| Notation | Notion équivalente pour un système en attente | Notion équivalente pour un système en fonctionnement continu |
|---|---|---|
| ω | Probabilité d'occurrence de chocs létaux à la sollicitation par unité de temps | Fréquence d'occurrence de chocs létaux |
| μ | Probabilité d'occurrence de chocs non létaux à la sollicitation par unité de temps | Fréquence d'occurrence de chocs non létaux |
| ρ | Probabilité conditionnelle de défaillance d'un élément en situation de choc non létal | Probabilité conditionnelle de défaillance d'un élément en situation de choc non létal |

**Table 2.** Notations particulières aux paramètres du modèle d'Atwood

## 3 Expressions et propriétés

Le modèle d'Atwood est défini au niveau d'un élément spécifique dans un groupe DCC. Les expressions du modèle sont :

Probabilité d'occurrence d'une défaillance d'un élément seul :

$$Q_1^{(N)} = Q_{IND} + \mu\rho(1\text{-}\rho)^{N-1} \quad \{1\}$$

Probabilité d'occurrence d'une défaillance d'un élément dans un groupe spécifique de taille N> k>1 :

$$Q_k^{(N)} = \mu\rho^k(1\text{-}\rho)^{N-k} \quad \{2\}$$

Probabilité d'occurrence d'une défaillance d'un élément dans un groupe de taille N:

$$Q_N^{(N)} = \mu\rho^N + \omega \quad \{3\}$$

Pour exprimer $Q_{TOT}$ au niveau d'un élément appartenant à un groupe de taille N, il faut prendre en compte le nombre de combinaisons possibles :

$$Q_k^{(N)} = \sum_{k=1}^{N} C_{k-1}^{N-1}\mu\rho^k(1-\rho)^{N-k} \quad \{4\}$$

[1] Cette notation est conventionnelle depuis les premiers rapports NUREG

[2] Parfois noté Qi ou Q1

Ainsi la probabilité de défaillance d'un élément spécifique Ei dans un groupe de N éléments, toutes causes confondues, s'exprime :

$$Q_{TOT} = Q_{IND} + \omega + \sum_{k=1}^{N} \binom{N-1}{k-1} \mu\rho^k (1-\rho)^{N-k} \qquad \{5\}$$

- Ce modèle distingue $Q_1^{(N)}$ et $\mathrm{Q_{IND}}$ : il considère comme possible des DCC affectant 1 élément (terme en $\mu\rho(1-\rho)^{N-1}$).
- Il attribue deux causes possibles à des DCC affectant l'ensemble du système ; un choc létal ou un choc non létal affectant par hasard l'ensemble des éléments du système.
- Il néglige les défaillances multiples indépendantes, combinées aux DCC partielles.
- On observe que : $\omega+\mu < 1$.
- Le terme $\omega$ permet de représenter les situations dans lesquelles les défaillances des éléments ne sont plus indépendantes, en considérant de façon conservative qu'elles peuvent cependant causer des DCC totales.

Avec une hypothèse de taux de défaillance faibles, il y a similarité des expressions mathématiques impliquant des probabilités ou des taux de défaillances lorsque les relations sont linéaires. Ainsi, pour un élément spécifique dans un groupe de N éléments, le taux de défaillance total s'exprime:

$$\lambda_{TOT} = \lambda_{IND} + \lambda_{\omega} + \sum_{k=1}^{N} \binom{N-1}{k-1} \lambda_{\mu}\rho^k (1-\rho)^{N-k} \qquad \{6\}$$

Il est possible d'appliquer le modèle d'Atwood aux systèmes en attente et aux systèmes en fonctionnement continu moyennant des adaptations des significations des paramètres.

Le modèle d'Atwood emploie les valeurs par défaut suivantes :

| Paramètre | Description | Valeurs par défaut |
|---|---|---|
| $\alpha = \frac{\lambda_{\mu}}{\lambda_{TOT}}$ (ou $\alpha = \frac{\mu}{Q_{TOT}}$) | Proportion de chocs non létaux (par rapport à l'ensemble des situations observées) | 0,405 |
| $\rho$ | Probabilité conditionnelle de défaillance d'un élément en situation de choc non létal | ρ = 0,2 ou ρ = 0,33. Pour des études de sensibilité, la valeur ρ = 0,5 est employée. |
| $\beta_{\mathrm{letal}} = \frac{\lambda_{\omega}}{\lambda_{\mathrm{TOT}}}$ (ou $\beta_{letal} = \frac{\omega}{Q_{TOT}}$) | Proportion de chocs létaux (par rapport à l'ensemble des situations observées) | $b_{lethal} = 5.10^{-3}$ |

**Table 3.** Valeurs pour la quantification du modèle d'Atwood

Ces termes permettent d'exprimer un Facteur Beta au niveau d'un élément Ei:

$$\beta_{Ei} = \frac{\omega + \sum_{k=1}^{N} \binom{N-1}{k-1} . Q_k^{(N)}}{Q_{TOT}} \qquad \{7\}$$

avec

$$Q_{TOT} = Q_{IND} + \omega + \sum_{k=1}^{N} \binom{N-1}{k-1} \mu\rho^k (1-\rho)^{N-k} \qquad \{8\}$$

Dans cette étude, nous proposons de déterminer les valeurs de ces paramètres par deux approches : une approche analytique et une, seconde, basée sur les résultats de simulations de Monte-Carlo.

Dans cet article, nous proposons également d'intégrer le modèle d'Atwood dans un modèle par réseaux de Petri (RdP) colorés d'un système numérique. Pour évaluer l'impact des DCC sur la disponibilité d'un système, le modèle d'Atwood est intégré dans un modèle complet du système, qui en représente le comportement dynamique (reconfigurations, réparations, logiques de votes).

## 4 Notion de DCC orientée

Dans le modèle d'Atwood généralisé, les composants d'un groupe DCC ont tous la même probabilité d'être affectés par un choc non létal. C'est une limite à son aptitude à représenter la phénoménologie de défaillances affectant les systèmes programmés. En effet, au sein d'un groupe DCC, il existe une diversité, voulue ou non, parmi les composants, qui n'est pas le reflet d'un hasard, mais de la structure du système. Par exemple, les éléments d'un groupe DCC situés dans une même division ont plus de chance d'être affectés par un effet thermique que les éléments du même groupe DCC situés dans une autre division. Pourtant, ils font partie d'un même groupe DCC et, selon le modèle, ils présentent tous le même risque d'être affectés par cet effet thermique, quelle que soit leur division. Pour cette étude, nous avons donc développé le concept de « DCC orientée » pour représenter ce phénomène.

Dans un système à N éléments, un choc non létal affecte les éléments avec une probabilité conditionnelle p. Le nombre attendu de composants concernés est le produit $N \cdot p$. L'ensemble des N éléments est alors divisé en deux sous-ensembles A et B, contenant respectivement a et b éléments, tels que $a + b = N$. Soit $x_A$ (respectivement $x_B$) la probabilité qu'un composant du groupe A (respectivement B) est affecté par choc un non létal. En raisonnant sur l'espérance mathématique, nous avons :

$$a.x_A + b.x_B = N.p \qquad \{9\}$$

Pour déterminer x et y, nous choisissons dans quelle proportion une DCC aura une incidence sur le groupe A (proportion $p_A$) ou sur le groupe B (proportion $p_B$), avec $p_A + p_B = 1$. Ainsi, on obtient :

$$x_A = \frac{N.p.p_A}{a} \qquad \{10\}$$
$$x_B = \frac{N.p.p_B}{b} \qquad \{11\}$$

Les solutions $x_A$ et $x_B$ de ce système sont les probabilités de choc non létal. Elles seront employées dans les jetons du RdP coloré qui représentent les éléments du groupe A et du groupe B.

## Architecture du système et hypothèses relatives à la sûreté de fonctionnement

La structure du système modélisé est la suivante.

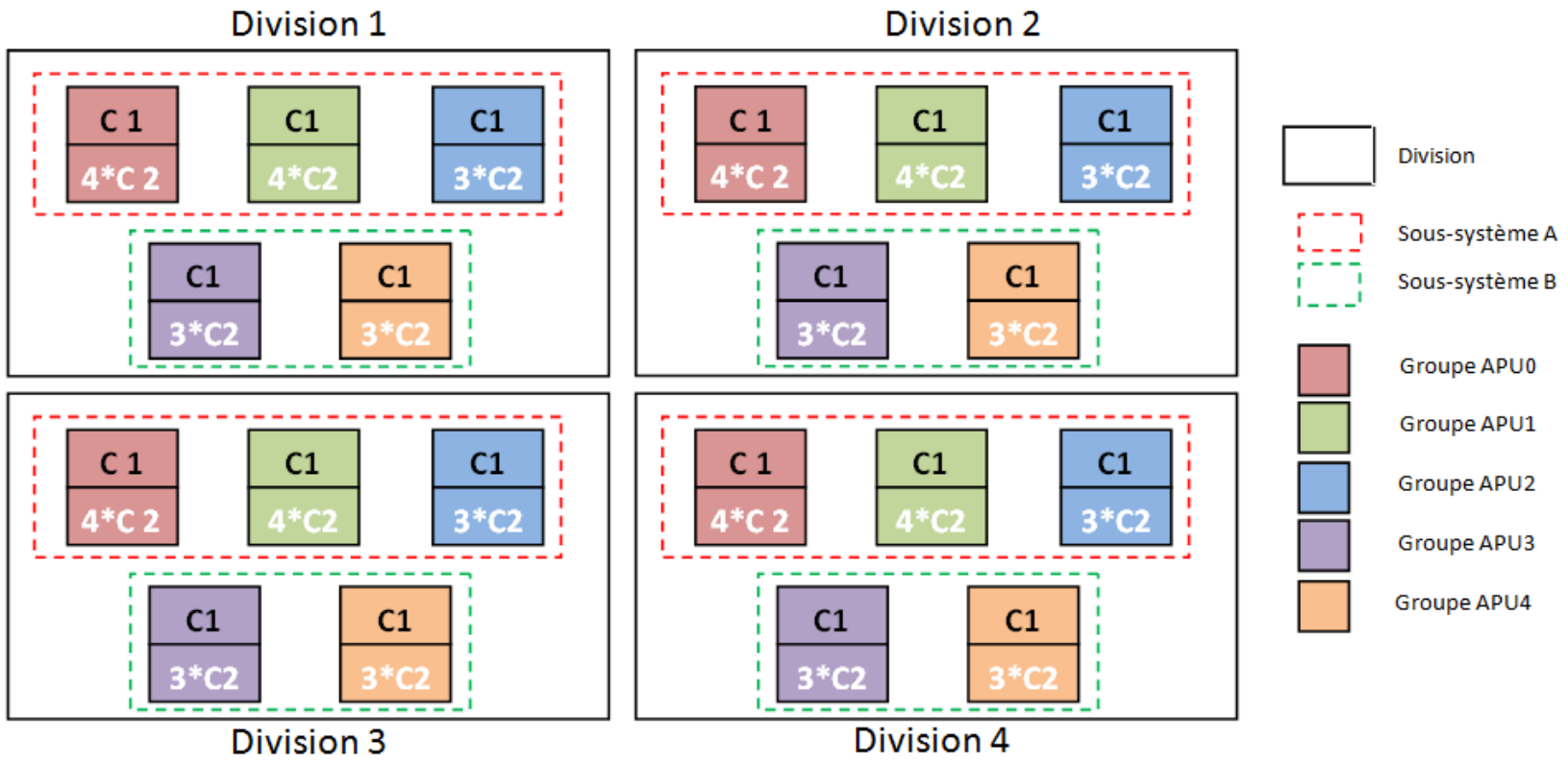

**Figure 1.** Architecture du système de contrôle commande modélisé

Ce système de protection comporte quatre divisions, qui sont strictement identiques. Ces divisions sont physiquement et électriquement séparées. Chaque division est composée de cinq unités d'acquisition et de traitement (APU). Les APU 0, 1 et 2 forment le sous-système A (SSA). Les APU 3 et 4 forment le sous-système B (SSB). Le regroupement des APU en sous-systèmes prend en compte la répartition des fonctions de contrôle du système d'I&C. Pour une même fonction, il y a une mise en œuvre dans un APU du SSB et une mise en œuvre dans une APU de SSB, avec différentes entrées et de traitement, sachant que leurs résultats doivent être identiques dans un mode de fonctionnement normal. Des cartes électroniques, C1 et C2, sont présentes dans chaque APU. Chaque APU contient une carte C1. Les APU 0 et 1 contiennent quatre cartes C2, et les APU 3, 4 et 5 contiennent trois cartes C2. Ces cartes électroniques sont utilisées pour l'acquisition, le traitement et l'émission de signaux. Sur cette base, un second type de partition est définie : les groupes d'APU (GAPU). Un groupe d'APU contient toutes les APUi (i de 0 à 4) des quatre divisions et constitue un groupe DCC.

### Hypothèses pour la modélisation du matériel électronique

Un taux de défaillance constant est supposé pour les cartes. Les trois cartes C2 d'une APU sont considérées comme un système série. Les défaillances des cartes peuvent être détectées par des autotests. Lorsqu'une défaillance est détectée par un autotest (défaillance type SA), le temps de détection est considérée comme nul. Si la défaillance n'est pas détectée par un autotest (défaillance type NSA), elle est détectée hors ligne pendant un test périodique. Pour une division donnée, ces tests périodiques ont lieu tous les18 mois. Ainsi, à chaque quart de cette période, une division est testée lors de l'essai périodique. Après cet essai, les cartes défectueuses sont réparées. Selon le fournisseur de cartes électroniques, la couverture des autotests des cartes C1 et C2 est de 100 %, mais cette valeur semble être peu conservative. Nous ajoutons donc des défaillances non détectées par autotest (NSA) afin de prendre en compte des erreurs dues à l'exploitation (erreurs de paramètres ou de configuration). Le taux de couverture est réduit à 85 %. Par contre, le taux de défaillance global des cartes est considéré comme inchangé. Ainsi :

$$\lambda_{IND} = \lambda_{SA} + \lambda_{NSA} \qquad \{12\}$$
$$\lambda_{SA} = \alpha \cdot \lambda_{IND} \qquad \{13\}$$
$$\lambda N_{SA} = (1 - \alpha) \cdot \lambda_{IND} \qquad \{14\}$$

### Hypothèses pour la modélisation du système

L'événement dangereux est la défaillance à la sollicitation du système de protection. La survenue de cet événement est déterminée par l'état de la logique de vote de l'APU :
- Une APU est défaillante lorsqu'au moins une carte C1 ou C2 est défaillante.
- Un groupe d'APU (GAPU) est défaillant lorsque 3 APU sur 4 sont défaillantes (logique en 3oo4).
- Un sous-système (SSA ou SSB) est défaillant lorsqu'un GAPU est défaillant.
- Le système est défaillant lorsqu'un sous-système est défaillant (cas d'une logique de vote en 1oo2) ou lorsque les deux sous-systèmes sont défaillants (cas d'une logique de vote en 2oo2).

La durée de la mission du système est de dix ans, ce système étant à l'arrêt seulement pendant les opérations de maintenance décennale. L'événement dangereux peut se produire pendant les dix années où le système est utilisé. Alors, le système devient indisponible. Le système passe après réparation d'un état indisponible à disponible sans être rénové (n'est donc pas « As Good As New »), certaines cartes électroniques peuvent être encore défaillantes lorsque le système redevient disponible.

## Démarche de modélisation par réseaux de Petri colorés, temporisés, hiérarchiques

### 1 Généralités

Parmi les outils prenant en compte les effets dynamiques, les Chaînes de Markov ou les Réseaux de Petri (RdP) sont des modèles pertinents. Le modèle du facteur Beta a été représenté dans les chaînes de Markov par [Lilleheier, 2008] et dans les réseaux de Petri classiques par [Signoret et al. 2012]. Cependant, le principal inconvénient de ces modèles est l'explosion combinatoire due à leur taille, lorsque le système modélisé est grand. Pour pallier cet inconvénient, nous proposons d'utiliser les RdP colorés, temporisés, hiérarchiques, qui sont une extension des RdP Généralisés.

Les RdP colorés [Jensen, 1997], [Jensen et al., 2009] constituent un formalisme de modélisation à événements discrets combinant les capacités des réseaux de Petri avec les capacités d'un langage de programmation de haut niveau. La principale différence entre un RdP classique et un RdP coloré est que les jetons peuvent avoir des couleurs différentes représentant des types de données (par exemple, booléens, entiers ou structure de données plus complexe). La couleur et la hiérarchie permettent de créer des modèles génériques pour des composants identiques. Ceci facilite la maîtrise du modèle et la représentation de la redondance. La couleur permet aussi de représenter la complexité du vote, de conserver l'information pour faciliter l'évaluation probabiliste et le traitement statistique des simulations. De plus, les RdP colorés temporisés ou stochastiques, sont très pratiques pour représenter les taux de défaillances et les temps de réparations.

Les RdP colorés temporisés hiérarchiques sont décrits formellement dans [Jensen et al. 2009], [Jensen, 1997], et leur application à un système numérique dans [Pinna et al. 2013].

#### Le modèle de haut niveau

Le modèle est réalisé de façon modulaire grâce à la capacité hiérarchique du RdP. Les DCC, le matériel, la logique de vote et l'état du système sont modélisés dans des sous-réseaux de Petri. Ils évoluent par l'envoi et la réception de messages modélisés par des jetons qui s'échangent entre les sous-réseaux, qui représentent l'arrivée d'une défaillance indépendante, d'une DCC, d'une réparation ou d'une maintenance sur une APU. Les messages diffusés mettent à jour de façon dynamique l'état du système. La PFD est ensuite estimée par les durées d'indisponibilité au niveau du système. Ainsi, le réseau de Petri de haut niveau représentant le système de contrôle-commande de la figure 2 est constitué de modules suivants :

- génération des DCC (encadré en bleu)
- modélisation du système de contrôle-commande (encadré en rouge)
- représentation de l'état du système (encadré en vert)

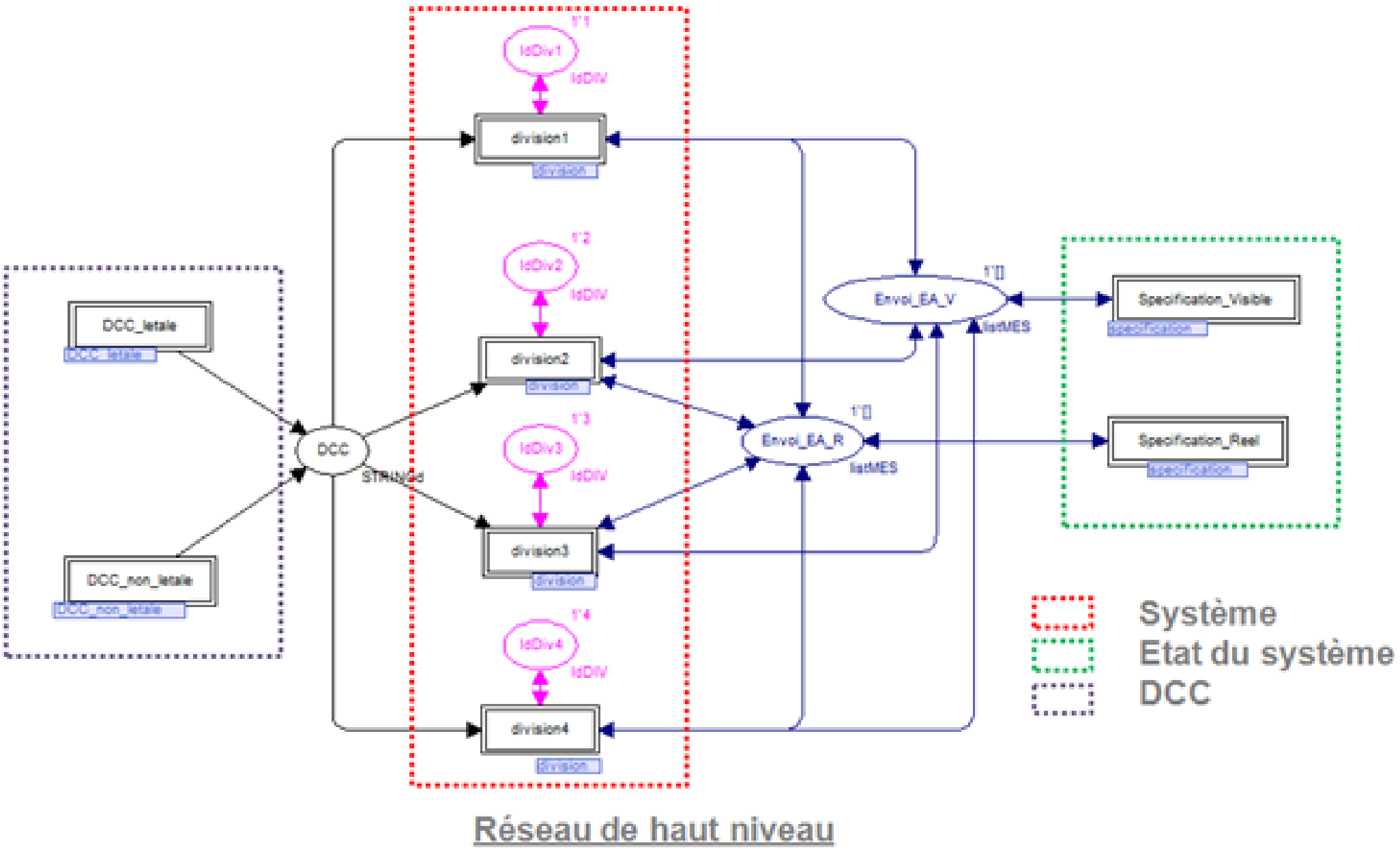

**Figure 2.** Réseau de Petri coloré de haut niveau représentant le système de contrôle-commande

### 2 Génération des DCC

**DCC non-létales.** Les DCC non-létales sont modélisés par le sous-réseau de Petri coloré de la figure 3 qui correspond à la transition de substitution « *DCC_non_letale* » de la figure 2. La place « *Nb-carte* » contient le nombre de cartes électroniques *N* du système. Le franchissement de la transition « *Save* » conserve le nombre de cartes *N* du système dans la place « *SNB_carte* » et place *N* jetons dans la place « *nb_carteu* ». La transition « *proba* » est ensuite franchie *N* fois et la fonction « *defdcc()* » effectue un tirage aléatoire de loi uniforme sur l'intervalle [0, 1] pour chaque carte. Si la valeur tirée est inférieure à la probabilité conditionnelle *p*, alors la carte considérée sera sensible au choc. Cette fonction retourne ainsi la valeur 0 (pour carte non-sensible au choc) ou 1 (carte sensible au choc).
Le franchissement de la transition « *init_temps* » permet de calculer l'instant de l'occurrence du choc non-létal en utilisant la fonction « *floor(exponential( !mu)+0.5)* » et, en même temps, de préciser si la défaillance est détectée par autotest ou non grâce à la fonction « *detect()* ». Cette fonction effectue in tirage aléatoire de loi uniforme sur l'intervalle [0, 1]. Si la valeur obtenue est inférieure au taux de couverture des essais périodiques $t_c$, le choc non-létal est détecté. Dans ce cas la fonction retourne la valeur 1, sinon elle retourne 0 pour le choc non-détecté.
Le franchissement de la transition « *dcc* » associe l'instant de l'occurrence de la DCC et la valeur de la variable de détection de la défaillance à chacun des jetons représentant les cartes sensibles au choc. Les cartes sensibles à la DCC seront choisies aléatoirement pendant la simulation du réseau de Petri.
Le franchissement de la transition « *no_def_dcc* » permet de supprimer les jetons représentant les cartes non sensibles au choc qui resteront en fonctionnement.
La transition « *new_dcc* » permet de déterminer l'instant de l'occurrence du nouveau choc non-létal et de redéfinir le nombre de cartes sensibles à ce nouveau choc.

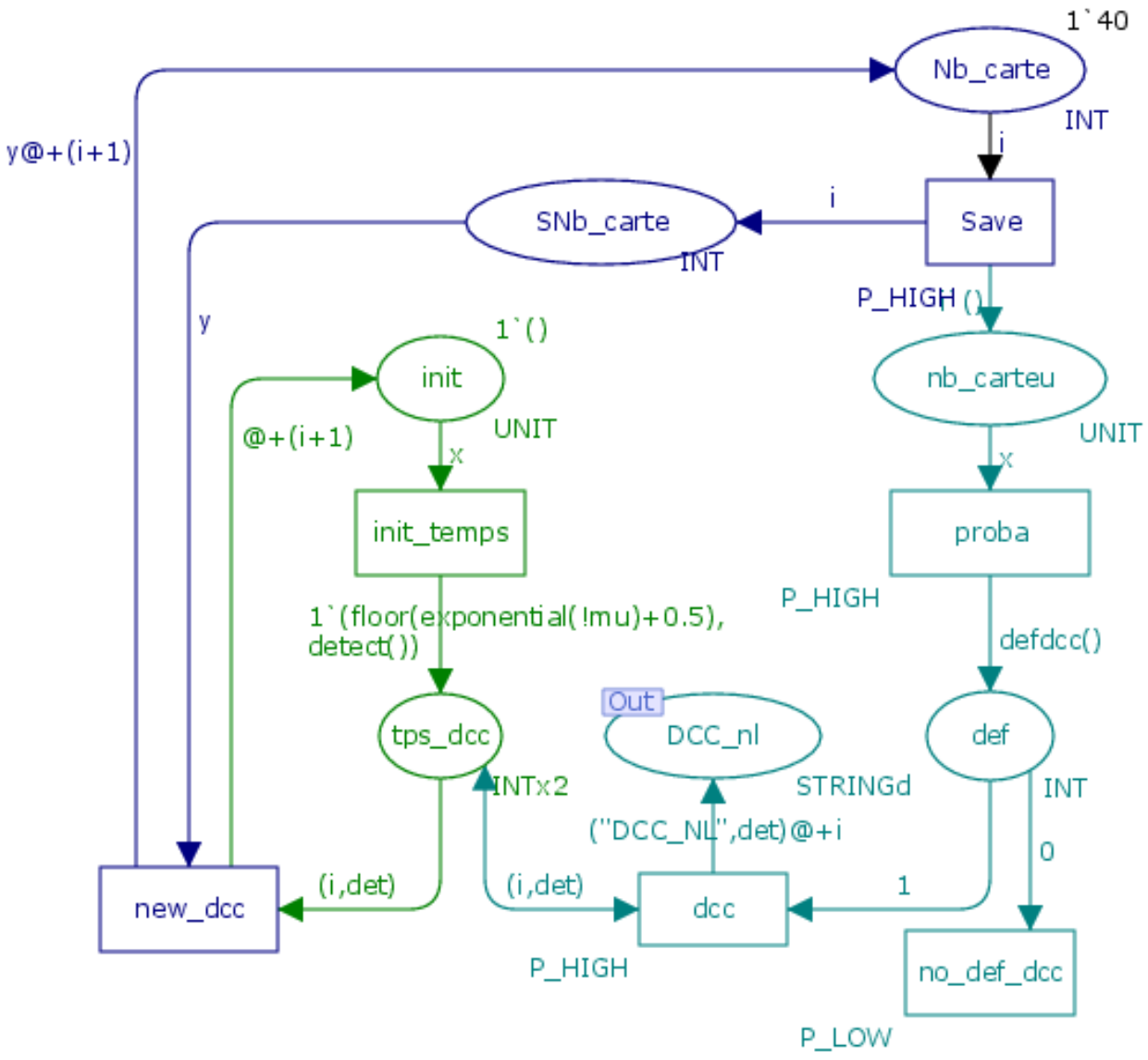

**Figure 3.** Modélisation des défaillances non-létales par réseau de Petri coloré

**DCC létales.** Les DCC létales sont modélisées par le sous-réseau de Petri coloré de la figure 4 qui correspond à la transition de substitution « *DCC_letale* » de la figure 2. Le franchissement de la transition « *gene_dcc_l* » permet de calculer l'instant de l'occurrence du choc létal par la fonction « *floor(exponential(!omega)+0.5)* ». Une DCC létale affecte tous les composants (toutes les cartes électroniques) du système et elle est détectée en ligne. Ainsi *N* jetons temporisés ayant la couleur "*DCC_L,1*" (« 1 » pour la détection) sont générés. Une nouvelle DCC létale est calculée une fois que la précédente ait lieu.

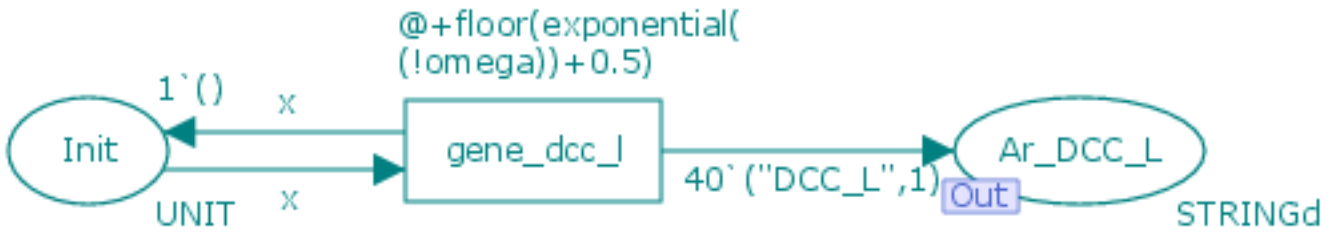

**Figure 4.** Modélisation des défaillances létales par réseau de Petri coloré

**Approche analytique pour l'estimation des paramètres du modèle d'Atwood.** Le modèle d'Atwood introduit trois paramètres à estimer $p$, $\mu$ et $\omega$. Deux approches, analytique ou par simulation, peuvent être utilisées pour réaliser leur estimation. Leur comparaison permet une vérification partielle de la validité de la modélisation. A partir des expressions du modèle d'Atwood (tables 2 et 3), nous obtenons le système d'équations suivant :

$$(1-\alpha\sum_{k=1}^{N}p^k(1-p)^{N-k})\mu+\alpha\omega=\alpha\lambda_{IND} \quad \{15\}$$

$$(-\beta\sum_{k=1}^{N}p^k(1-p)^{N-k})\mu+(1-\beta)\omega=\beta\lambda_{IND} \quad \{16\}$$

La valeur de probabilité conditionnelle $p$ est considérée comme connue (valeurs par défaut classiques de 0,2, 0,33 ou 0,5). Le taux de défaillances indépendantes pour une carte électronique est fournie par le constructeur : $\lambda_{IND}$ = 2,35 $10^{-6}$ $h^{-1}$. Les équations {15} et {16} donnent les valeurs de fréquences d'occurrence des chocs non létaux et létaux. Les résultats obtenus par résolution du système sont présentés dans la table 4.

| p | $\mu\ in\ h^{-1}$ | $\omega\ in\ h^{-1}$ |
|---|---|---|
| 0.2 | $9.52\ 10^{-7}$ | $1.18\ 10^{-8}$ |
| 0.33 | $9.52\ 10^{-7}$ | $1.18\ 10^{-8}$ |
| 0.5 | $9.52\ 10^{-7}$ | $1.18\ 10^{-8}$ |

**Table 4:** Valeurs des fréquences d'occurrence des chocs non létaux et létaux en fonction de p par approche analytique

**Approche par simulation de Monte-Carlo pour l'estimation des paramètres du modèle d'Atwood.** Nous introduisons le rapport entre le taux de défaillances indépendantes et le taux de défaillance total d'une carte électronique :

$$\gamma = \frac{\lambda_{IND}}{\lambda_{TOT}} = 1 - \alpha.\sum_{k=1}^{N} C_{N-1}^{k-1} p^k.(1-p)^{N-k} - \beta_{letal} \qquad \{17\}$$

$$\gamma = 1 - \alpha.p - \beta_{letal} \qquad \{18\}$$

Nous introduisons également $E_i$, qui représente le nombre de défaillances indépendantes observées après des simulations de missions de 10 ans.

$$\lambda_{IND} \;\; = \;\; \frac{E_i}{10\ ans\ * N} \qquad \{19\}$$

Ce nombre est obtenu en simulant seulement les défaillances indépendantes dans le modèle réseau de Petri et en dénombrant leurs occurrences. Il permet d'obtenir, en utilisant l'équation {19}, le taux de défaillances indépendantes $\lambda_{IND}$ d'une carte dans le système. Les autres taux peuvent être estimés à partir des équations suivantes :

$$\mu \;\; = \;\; \frac{\alpha.\lambda_{IND}}{\gamma} \qquad \{20\}$$

$$\omega \;\; = \;\; \frac{\beta_{letal}.\lambda_{IND}}{\gamma} \qquad \{21\}$$

La valeur attendue de $E_i$ est égale à 8,28 et la valeur obtenue pour le taux de défaillances indépendantes est $\lambda_{IND} = 2{,}36.10^{-6}\ h^{-1}$. Les résultats obtenus par l'application de cette approche sont présentés dans la table 5.

| p | $\gamma$ | $\lambda_{TOT}$ | $\mu$ | $\omega$ |
|---|---|---|---|---|
| 0.2 | 0.995 | $2.37\ 10^{-6}$ | $9.62\ 10^{-7}$ | $1.18\ 10^{-8}$ |
| 0.33 | 0.995 | $2.37\ 10^{-6}$ | $9.62\ 10^{-7}$ | $1.18\ 10^{-8}$ |
| 0.5 | 0.995 | $2.37\ 10^{-6}$ | $9.62\ 10^{-7}$ | $1.18\ 10^{-8}$ |

**Table 5**: Valeurs des fréquences d'occurrence des chocs non létaux et létaux en fonction de p par approche de Monte Carlo

Les fréquences d'occurrence des chocs non létaux et létaux obtenues par les deux approches (approche analytiques et simulation) n'ont pas d'écart significatif. Ce résultat permet de valider la représentation du modèle d'Atwood par RdP. Le REX n'est pas suffisant pour valider l'approche analytique.

## 5 Modélisation des APU et des cartes

Les aspects hiérarchiques et modulaires des RdP colorés sont exploités pour développer le modèle du système de contrôle-commande. Ainsi, chaque division est modélisée par une transition de substitution dans le modèle de haut niveau du système de la figure 2. Une division est constituée de cinq APU, chaque APU contenant le nombre correct des cartes électroniques $C_1$ et $C_2$. Les défaillances de cause commune sont transmises aux APU et aux cartes électroniques par l'intermédiaire des places du réseau de Petri. L'état de l'APU est déterminé en fonction de l'état de ses cartes (disponibles ou indisponibles). Dès que l'état des cartes change, le nouvel état de l'APU est transmis à la partie spécification (encadré vert de la figure 2) qui déterminera l'état du système global.

Une carte électronique, représentée sur la figure 5, a trois états possibles : opérationnelle (place « *Marche* »), défaillante et non-détectée (place « *detection* ») et en réparation (place « *réparation* »). Lorsque la carte est opérationnelle, son état peut changer suite à l'occurrence de l'un de quatre événements suivants (les quatre transitions en sortie de la place « *Marche* ») :

- Défaillance indépendante détectée en ligne par autotest (générée sur la partie mauve à droite du RdP)
- Défaillance indépendante détectée hors ligne par essai périodique (générée sur la partie mauve à gauche du RdP)
- DCC non-létale détectée en ligne ou DCC létale (modélisées par les jetons reçus dans la place « *AR_dcc* » depuis les modèles des DCC présentés dans les figures 3 et 4)
- DCC non-létale détectée hors-ligne par essai périodique (modélisée également par les jetons reçus dans la place « *AR_dcc* » depuis le modèle des DCC présenté dans la figure 3).

Si les défaillances sont détectées par essai périodique, la transition « *rep* » est franchie et la carte change d'état pour être réparée (place « *réparation* »). Si les défaillances sont détectées en ligne, la carte change d'état directement depuis l'état opérationnel (place « *Marche* ») vers l'état en réparation. Après la réparation de la carte (transition « *rep_ok* »), elle redevient opérationnelle. L'instant de la prochaine défaillance indépendante détectée en ligne ou hors-ligne est également calculé.

Le modèle réseau de Petri coloré de la figure 5 est générique pour toutes les cartes électroniques, seulement les valeurs numériques des paramètres (taux de défaillance et de réparation) sont différentes pour les cartes $C_1$ et $C_2$.

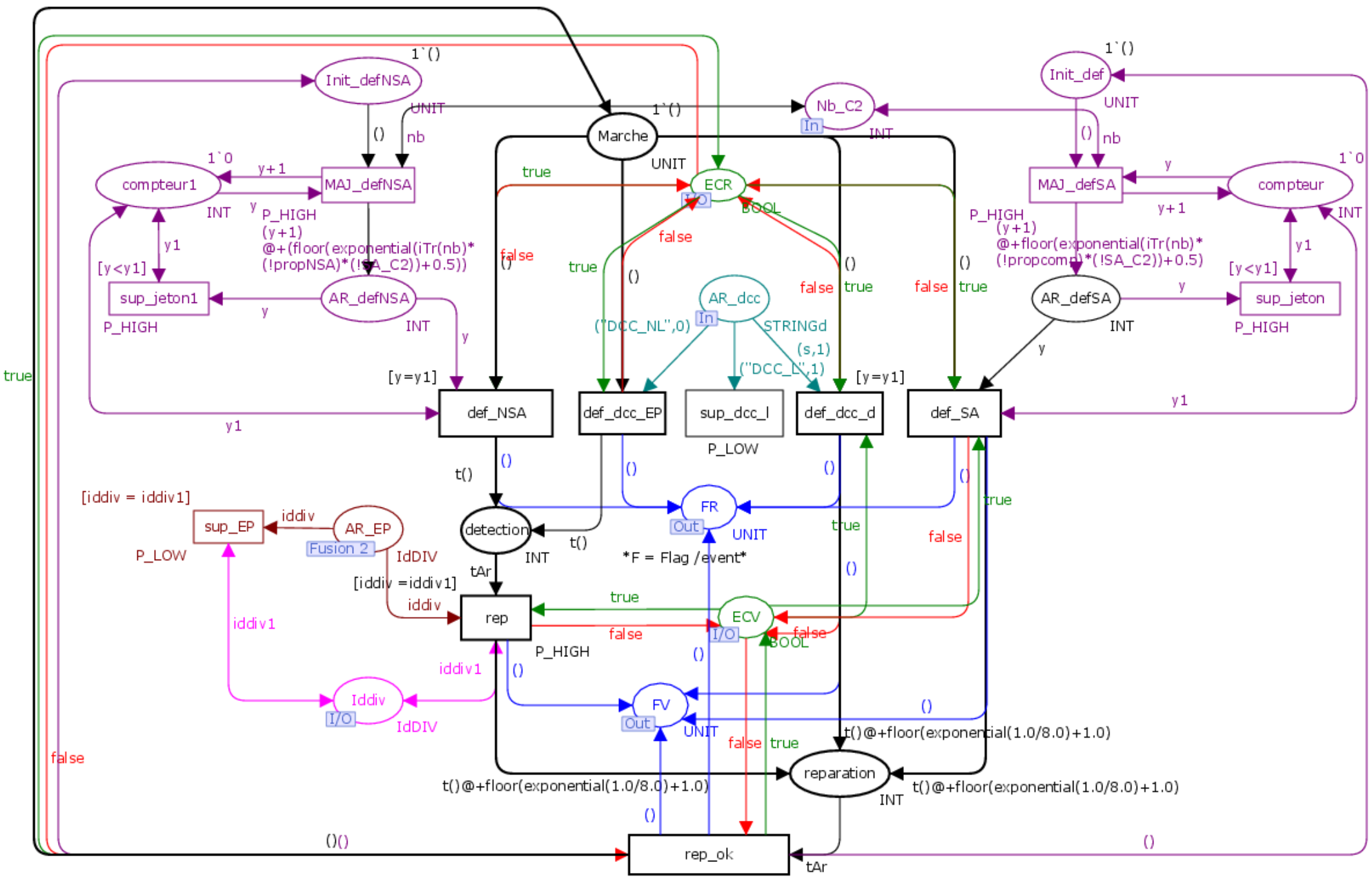

**Figure 5.** Réseau de Petri coloré représentant le comportement fonctionnel et dysfonctionnel d'une carte électronique

## 6 Modélisation de l'état du système

A partir des informations concernant l'état (disponible/indisponible) des APU $i$ ($i \in [0, 4]$ de la division $j$, $j \in [1, 4]$), il est possible de déterminer l'état du système pendant la simulation de Monte-Carlo du réseau de Petri de la figure 6. L'état du système est représenté par un jeton dont la couleur est constituée de cinq variables booléennes, chacune représentant l'état des cinq APU. Suivant la couleur de ce jeton, le système est disponible ou indisponible. Les différentes configurations exprimées par les conditions de garde des transitions permettent de définir les conditions de disponibilité et d'indisponibilité du système (ici pour une logique en 2oo2).

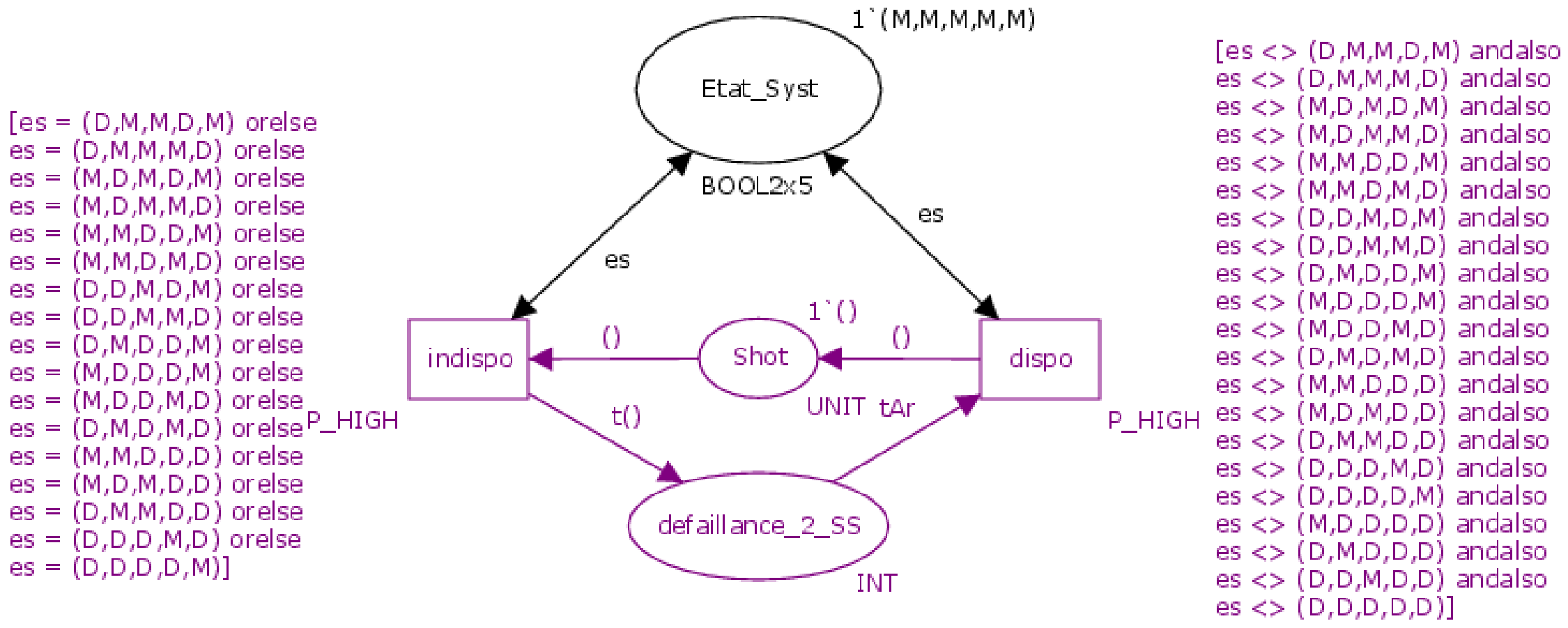

**Figure 6.** Réseau de Petri coloré permettant de déterminer l'état du système global (disponible / indisponible)

Le réseau de Petri obtenu pour la modélisation du système de contrôle-commande et de la modélisation et la propagation des DCC suivant le modèle d'Atwood est formé de 504 places et 685 transitions. Même si la taille de ce modèle est assez grande, les concepts de *hiérarchie* et des *couleurs* ont permis d'obtenir, par instanciation des sous-modèles génériques, *un modèle modulaire et lisible*. Nous pouvons noter qu'un réseau de Petri classique (sans couleurs et sans hiérarchie) équivalent pour le même système aura plusieurs milliers des places et transitions. Dans la section suivante, le modèle réseau de Petri obtenu est utilisé pour l'évaluation probabiliste des indicateurs de la sûreté de fonctionnement du système étudié.

# Résultats

## 1 Probabilité de défaillance à la sollicitation

Le système étudié dans cet article est un système de contrôle-commande dont sa mission est la protection d'une centrale nucléaire. Nous allons nous intéresser à l'évaluation de la probabilité de défaillance à la sollicitation (PFD - Probability of Failure on Demand) conformément à la norme IEC 61508 qui traite de la sécurité fonctionnelle des systèmes électroniques programmables. Nous définissons deux types de PFD :

- la *PFD réelle* ou *de référence* qui représente l'indisponibilité réelle du système même si les défaillances ne sont pas détectées ;
- la *PFD visible* qui représente l'indisponibilité du système à partir du moment où les défaillances sont détectées.

La définition de ces deux types de PFD permet d'évaluer la différence entre l'état réel du système et l'état perçu par les opérateurs d'une centrale pendant l'exploitation.
La PFD est évaluée par la simulation de Monte-Carlo du réseau de Petri coloré en utilisant l'équation suivante :

$$PFD = \frac{Durée\ pendant\ laquelle\ le\ système\ est\ indisponible}{Durée\ totale\ de\ la\ mission\ du\ système} \quad \{12\}$$

Les résultats obtenus par la simulation de 10000 histoires pendant une période de 10 ans (entre deux visites décennales d'une centrale nucléaire) sont présentés dans le tableau 3. L'intervalle de confiance à 95% pour les résultats présentés dans le tableau 3 a été également estimé.

| **Mesure** | $p$ | **Valeur moyenne** |
|---|---|---|
| *PFD référence* | 0.2 | $2.5 \cdot 10^{-5}$ |
| *PFD visible* | 0.2 | $< 10^{-6}$ |
| *PFD référence* | 0.33 | $2.73 \cdot 10^{-4}$ |
| *PFD visible* | 0.33 | $1.0 \cdot 10^{-6}$ |
| *PFD référence* | 0.5 | $5.4 \cdot 10^{-4}$ |
| *PFD visible* | 0.5 | $4.0 \cdot 10^{-6}$ |

**Tableau 3.** PFD référence et visible du système de contrôle-commande pour différentes valeurs de $p$

On constate que si la probabilité conditionnelle de défaillance d'un composant en situation de choc non-létal $p$ augmente, il y aura plus de composants affectés par le choc et la PFD augmente également. L'écart entre la PFD référence et la PFD visible est de l'ordre de 10 à 100 (fortes valeurs de ).

## 2 Influence de l'orientation des DCC non-létales

Afin d'évaluer l'impact de l'orientation des DCC non-létales, on introduit une propagation asymétrique des DCC visant les deux sous-systèmes SSA et SSB.
La probabilité conditionnelle de défaillance d'un composant en situation de choc non-létal est $p = 0.2$. La fréquence des chocs non-létaux est fixée arbitrairement à un choc par an, soit $\mu = 1.14 \cdot 10^{-1}$. Les chocs létaux n'ayant aucune influence sur la propagation asymétrique des DCC ne sont pas représentés. 10000 histoires d'une durée de 10 ans chacune sont considérées pour la simulation de Monte-Carlo et la simulation d'une histoire est arrêtée lorsque le système est défaillant. Le tableau 4 présente les résultats obtenus. La somme des combinaisons des défaillances (colonnes du tableau) est égale au nombre total d'histoires (10000) pour une orientation de DCC considérée.

| **Orientation DCC** | $\mathbf{0.1}$<br>$\mathbf{0.9}$ | $\mathbf{0.2}$<br>$\mathbf{0.8}$ | $\mathbf{0.3}$<br>$\mathbf{0.7}$ | $\mathbf{0.4}$<br>$\mathbf{0.6}$ |
|---|---|---|---|---|
| DCC suffis. | 9404 | 9723 | 9787 | 9823 |
| C. déf SA | 3 | 0 | 1 | 1 |
| C. déf NSA | 21 | 9 | 9 | 6 |
| C. déf SA et NSA | 77 | 46 | 42 | 37 |
| C. DCC et déf SA | 8 | 6 | 1 | 1 |
| C. DCC et déf NSA | 487 | 216 | 160 | 132 |
| MTTFF relatif [h] | 1 | 0,58 | 0,44 | 0,39 |

**Tableau 4.** Combinaisons des cartes défaillantes menant à l'indisponibilité du système en fonction de $p_1$ et $p_2$ (quatre cas d'asymétrie).

Plus un sous-système est privilégié (en fonction de $p_1$ et $p_2$), plus le MTTFF augmente. Le rôle de la logique de vote explique ce phénomène. Nous observons aussi que les combinaisons de défaillances indépendantes avec ou sans DCC mènent rarement à la défaillance du système, dû au niveau élevé de la redondance. Négliger ce phénomène dans les expressions analytiques pourrait ainsi être justifié dans certains cas. Les défaillances indépendantes détectées hors ligne mènent plus facilement à la défaillance du système que celle détectées en ligne. Réduire la périodicité des essais périodiques ou augmenter le taux de couverture des cartes sont deux moyens pour réduire ce phénomène.

## Conclusion

Dans cet article, les réseaux de Petri colorés ont été utilisés pour modéliser des systèmes programmés de contrôle-commande de grande taille et évaluer leurs indicateurs de sûreté de fonctionnement en présence des défaillances de cause commune (DCC). Le modèle d'Atwood a été considéré pour modéliser les DCC et il a été représenté par des réseaux de Petri. Ce modèle permet la prise en compte des combinaisons de défaillances indépendantes, de DCC létales, et de DCC non-létales. Une extension de ce modèle a été proposée afin de représenter la propagation asymétrique des défaillances non-létales affectant préférentiellement certaines parties du système. Cette extension permet de relâcher l'hypothèse qui concerne la répartition uniforme des chocs non-létaux sur l'ensemble des composants et de prendre en compte l'effet de la diversité entre parties du système. Cette approche de modélisation et d'évaluation de la sûreté de fonctionnement a été appliquée à un système réel de contrôle-commande de protection d'une centrale nucléaire.

Le très faible retour d'expérience sur ces systèmes de contrôle-commande rend difficile l'estimation des paramètres probabilistes qui représentent les DCC. Nous nous proposons pour la suite d'utiliser ce modèle par réseaux de Petri colorés pour constituer des retours d'expérience simulés en conservant tous les événements produits sur une période donnée qui permettront ensuite une estimation plus efficace des facteurs probabilistes (Facteurs Beta et Facteurs Alpha).

Cette approche par simulation de Monte-Carlo pourra être utilisée pour valider les hypothèses induites par la loi binomiale pour représenter les défaillances de cause commune.

## Références